\documentclass[letterpaper,12pt,titlepage]{article}
\usepackage{a4wide}
\usepackage{amsfonts}
\usepackage{amsmath}
\usepackage{graphicx}
\usepackage{latexsym}
\numberwithin{equation}{section}

\immediate\write18{./bibgen \jobname}
\usepackage{youngtab}\Yvcentermath1

\def\fund{  \> {\vcenter  {\vbox  
              {\hrule height.6pt
               \hbox {\vrule width.6pt  height5pt  
                      \kern5pt 
                      \vrule width.6pt  height5pt }
               \hrule height.6pt}
                         }
                   }
           \>\> }

\def\antifund{  \> \overline{ {\vcenter  {\vbox  
              {\hrule height.6pt
               \hbox {\vrule width.6pt  height5pt  
                      \kern5pt 
                      \vrule width.6pt  height5pt }
               \hrule height.6pt}
                         }
                   } }
           \>\> }

\begin{document}

\begin{titlepage}
\begin{flushright}
UT-06-27\\
hep-th/0612278
\end{flushright}

\vskip 1cm

\begin{center}
\textbf{\LARGE
Quantum moduli space of the cascading $Sp (p+M)\times Sp(p)$ gauge theory
}

\vskip1.5cm

{\Large Fumikazu Koyama and Futoshi Yagi}\\

\vskip1cm

\textit { { Department of Physics, Faculty of Science,} \\
{University of Tokyo, Tokyo 113-0033,  Japan}}\\

\vskip1cm

{\large\textbf{Abstract}}

\end{center}
We extend the detailed analysis of the quantum moduli space of the 
cascading $SU(p+M)\times{}SU(p)$ gauge theory in the recent 
paper of Dymarsky, Klebanov, and Seiberg for the $Sp(p+M)\times{}Sp(p)$ 
cascading gauge theory, which lives on the world volume of 
$p$ D3-branes and $M$ fractional D3-branes at the tip of the 
orientifolded conifold. 
As in their paper, we also find in this case 
that the ratio of the deformation parameters of the quantum constraint 
on the different branches in the gauge theory can be reproduced by the ratio 
of the deformation parameters of the conifold with different numbers 
of mobile D3-branes.

\vbox{}\vspace{1\fill}

\end{titlepage}

\section{Introduction}
In the recent paper \cite{DKS}, Dymarsky, Klebanov, and Seiberg gave a 
detailed analysis of the moduli space of the cascading $SU(p+M)\times{}SU(p)$
gauge theory, which lives on the world volume of $p$ D3-branes and 
$M$ fractional D3-branes at the tip of the conifold \cite{KS}. 
They found new mesonic branches, which is characterized by the number 
of performing the cascading duality transformation and the phase factor 
of the deformation parameter. The latter can also be seen by the 
$\mathbb{Z}_{2M}$ 
R-symmetry breaking by gaugino condensation to $\mathbb{Z}_2$. 
They discussed that the branches correspond to the warped deformed conifold 
with different numbers of mobile D3-branes. In particular, they showed that 
the ratio of the deformation parameters of the quantum constraints 
on the different branches in the field theory 
can be reproduced by the ratio of the deformation parameters 
of the conifold with different numbers of mobile D3-branes.

In this note, we will extend their analysis for the cascading 
$Sp(p+M)\times{}Sp(p)$ gauge theory \footnote{In our notation, 
$Sp(1)\simeq{}SU(2)$.}\cite{NSW}, which is dual 
to the string background given by 
$p$ D3-branes and $M$ fractional D3-branes at the tip of the 
orientifolded conifold with four D7-branes on top of an orientifold 
plane (O7-plane) to cancel the RR charge of the O7-plane. 
As discussed in \cite{NSW,SW}, the supergravity solution is given by 
the Klebanov-Strassler solution \cite{KS} only projected 
by the $\mathbb{Z}_2$ operation. 
Therefore, it is obvious that the ratio of the deformation parameters of 
the orientifolded conifold is identical to the one of the conifold in 
\cite{DKS}. In this note, we demonstrate that the cascading 
$Sp(p+M)\times{}Sp(p)$ gauge theory gives the similar result on the 
ratio of the deformation parameters to the one of 
the $SU(p+M)\times{}SU(p)$ gauge theory in \cite{DKS}.
This note is organized as follows: in the next section, 
we will give a brief review on the result of \cite{DKS}, 
in particular on the ratio of the deformation parameters. 
In the section 3, we will study the moduli space of the 
cascading $Sp(p+M)\times{}Sp(p)$ gauge theory and demonstrate 
that the ratio of the deformation parameters gives the correct ratio 
given on the string side in section 2.
The quantum moduli space for the case $p=1$ 
has been explored in the paper \cite{NSW}. 
Since no new mesonic branches are however available in the case, 
we need to consider other cases with $p>1$. 
The section 4 will be devoted to summary and discussion.
In the appendix, the quantum moduli space is discussed for 
$p=M-1$, $M$, which could happen at the last step of the cascading flow. 


\section{The Klebanov-Strassler Solution with Mobile D3-Branes}

It has been discussed in the paper \cite{NSW,SW} that 
the Klebanov-Strassler (KS) solution \cite{KS} projected by the $\mathbb{Z}_2$ 
projection is dual to $Sp(p+M)\times{}Sp(p)$ gauge theory. The solution is 
the warped deformed conifold background 
\begin{align}
ds^2_{10}=h^{-1/2}(r)dx^{\mu}dx_{\mu}+h^{1/2}(r)ds^2_6, \notag
\end{align}
where $h(r)$ is a warp factor and $ds_6^2$ is given by the metric of 
the deformed conifold \cite{CdelO}. For large $r$, the metric can 
be approximated by the singular conifold with the metric 
\begin{align}
ds_6^2\sim dr^2+r^2ds^2_{T^{1,1}}. 
\end{align}
Here, $T^{1,1}$ can be described as a $U(1)$ bundle over $S^2\times{S}^2$ 
\cite{CdelO,PP}. In this note, we follow the notation in \cite{LEC}, where 
the two two-spheres are parametrized by $(\theta_{1},\phi_{1})$ and 
$(\theta_{2},\phi_{2})$, and the $U(1)$ fiber by $\psi\in[0,4\pi]$. 

The deformed conifold can also be described by the polynomial 
\begin{align}
xy-zw=\varepsilon^2
\end{align}
of the complex variables $x,y,z,w$ in $\mathbb{C}^4$ 
with the deformation parameter $\varepsilon$.

In order to obtain the dual description of 
the cascading $Sp(p+M)\times{}Sp(p)$ 
gauge theory, we need to orientifold the warped deformed conifold by 
the $\mathbb{Z}_2$ projection 
\begin{align}
z \quad\leftrightarrow\quad w,
\end{align}
which is equivalent to 
\begin{align}
\left(\theta_1,\phi_1\right) \quad\leftrightarrow\quad 
\left(\theta_2,\phi_2\right). 
\label{projZ2}
\end{align}
Since the $\mathbb{Z}_2$ projection has the set of fixed points 
\begin{align}
xy-z^2=\varepsilon^2
\end{align}
and gives rise to an orientifold plane (O7-plane), we need to introduce 
four D7-branes to cancel the RR-charge of the O7-plane, otherwise 
not only the D3-brane charge, but also the D5-brane charge would change, 
as we go down to the tip of the conifold \cite{ouyang}. 
The four D7-branes will have the effect on the gauge theory side 
that additional fundamental matters are introduced in the gauge theory.  

On the gauge theory side, we begin with 
$SU(2M+2p)\times SU(2p)$ gauge theory which includes 
two bifundamentals $A_{1,2}$ in $(\tiny\yng(1),\overline{\tiny\yng(1)})$ 
and two conjugates $B_{1,2}$ in $(\overline{\tiny\yng(1)},\tiny\yng(1))$ 
and use the identification 
\begin{align}
\text{tr}[A_1B_2]\sim z, \qquad \text{tr}[A_2B_1]\sim w
\end{align}
to perform the corresponding $\mathbb{Z}_2$ projection 
in the field theory to give 
the $Sp(p+M)\times{}Sp(p)$ gauge theory \cite{NSW}. 
Therefore, it is natural to begin with twice as many of 
the D3-brane and D5-brane charge as in the $SU(p+M)\times{}SU(p)$ theory;
\begin{align}
\frac{1}{16\pi^4}\int_{T^{1,1}}F_5=2p, \qquad\frac{1}{4\pi^2}\int_{A}F_3=2M, 
\label{bc}
\end{align}
where $A$ is a copy of $S^3$ given by $(\theta_2,\phi_2)=(0,0)$.
Here, the D3-brane charge $2p$ is defined on the $T^{1,1}$ 
at the cut-off radius $r=r_c$. 

In order to obtain the ratio of the deformation parameters with 
the different numbers of mobile D3-branes, we only need to know 
the large radius $r$ behavior of the warped deformed conifold. 
In particular, we are interested in the self-dual five-form $F_5$ 
\begin{align}
F_5=\widetilde{F}_5+*\widetilde{F}_5,\qquad 
\widetilde{F}_5= 27\pi\mathcal{N}(r){\rm Vol}\left({T^{1,1}}\right), 
\end{align}
of the KS solution \cite{KS}. 
Here $\mathcal{N}(r)$ stands for the effective number of the 5-form flux 
through the deformed $T^{1,1}$ at $r$, and its radial dependence reflects 
the gravitational manifestation of the duality cascade. 
We can read off its large $r$ limit 
\begin{align}
\mathcal{N}(r)
\quad\rightarrow\quad 
\frac{g_s(2M)^2}{2\pi}\ln\left(\frac{r^3}{\varepsilon^2}\right)+{\rm const.}, 
\label{flux}
\end{align}
where ${\rm const.}$ denotes a constant which is independent of $r$ and 
$\varepsilon$. 

Now, we are ready to find the ratio of the deformation parameters, 
one of which is given by the KS solution with $(p-lM)$ mobile D3-branes 
and their mirrors at $r=r_{\text{D3}}$, where $r_{\text{D3}}$ satisfies 
$\varepsilon^{2/3}\ll{r}_{\text{D3}}<r_c$,
and the other of which given by the one with no mobile D3-branes. 
The condition for $r_{\text{D3}}$ means that $(p-lM)$ mobile D3-branes 
and their mirrors stay far from the tip of the conifold. 
Furthermore, for simplicity, we also suppose that 
all $(p-lM)$ mobile D3-branes and their mirrors 
lie at the same surface $r=r_{\text{D3}}$. 
We will see that our following analysis is independent of the choice of 
$r_{\text{D3}}$. 
In fact, one can verify that the result does not change, even if we place 
the D3-branes separately. We describe the geometries in the regions  
$r<r_{\text{D3}}$ and $r>r_{\text{D3}}$ by the KS solutions with 
the deformation parameters $\varepsilon_{c}$ 
and $\varepsilon_l$, respectively. 
As a matching condition, due to the presence of $(p-lM)$ D3-branes and 
their mirrors at $r_{\text{D3}}$, we can see that the number of 
the five-form flux jumps by $2(p-lM)$ at $r_{\text{D3}}$. 

We suppose that 
in the region of $r_{\text{D3}}<r\le r_c$ 
the effective number of the 5-form flux 
is given by (\ref{flux}) 
with the deformation parameter $\varepsilon=\varepsilon_c$, 
which we denotes $\mathcal{N}_+(r)$.  
The boundary condition  at the surface $r=r_c$ (\ref{bc}) becomes 
\begin{align}
2p=\frac{g_s(2M)^2}{2\pi}\ln\left(\frac{r_c^3}{\varepsilon_c^2}\right)
+{\rm const.}  
\label{prc}
\end{align} 
This equation determines the $\varepsilon_c$. 
After going down to the surface at $r=r_{\text{D3}}$, 
the number of 5-form flux becomes 
\begin{align}
\mathcal{N}_{+}(r_{\text{D3}})=\frac{g_s(2M)^2}{2\pi}\ln
\left(\frac{r_{\text{D3}}^3}{\varepsilon_c^2}\right)+{\rm const.} 
\label{NrD3+}
\end{align}

Below the surface $r=r_{\text{D3}}$, 
we use 
(\ref{flux}) with the deformation parameter 
$\varepsilon=\varepsilon_c$, 
which we denote $\mathcal{N}_-(r)$. 
We must take into account the jump of the five-form flux 
at $r=r_{D3}$, and the matching condition is given by
\begin{align}
\mathcal{N}_{+}(r_{\text{D3}})=\mathcal{N}_{-}(r_{\text{D3}})+2(p-lM),
\end{align}
Thus, the deformation parameter for the branch $l$ is 
\begin{align}
\varepsilon_l^2=\varepsilon_c^2{e}^{{2\pi\over{g}_s}{2p\over(2M)^2}}
\exp\left({-{2\pi{l}\over{}g_s(2M)}}\right).
\end{align}
We note that the deformation parameter is 
independent of $r_{\text{D3}}$, 
as mentioned above. 

Finally, we obtain the ratio between the deformation parameter 
$\varepsilon_l$ 
with $(p-lM)$ pairs of mobile D3-brane and its mirror and 
$\varepsilon_{l+1}$ with $(p-M-lM)$ ones as 
\begin{align}
\left(\frac{\varepsilon_{l+1}}{\varepsilon_l}\right)^{\frac{2}{3}}
=\exp\left(-\frac{2\pi}{3g_s(2M)}\right). 
\label{stringresult}
\end{align}
In the following sections, we will follow the proposal in \cite{DKS} that 
the meson branch after performing the cascade duality transformation $l$ 
times corresponds to the KS solution with $(p-Ml)$ mobile D3-branes and 
their mirrors. 

And using the relation explained later between the string coupling constant 
$g_s$ on the string side and the dynamical scale 
in the gauge theory, we will find that the ratio of the deformation parameters 
 on the different meson branches is in agreement with the above 
result on the ratio of the deformation parameters also in 
the $Sp(p+M)\times{}Sp(p)$ gauge theory, 
not only in the $SU(p+M)\times{}SU(p)$ gauge theory.


\section{The Moduli Space of the $Sp(p+M)\times{}Sp(p)$ Gauge Theory}

We consider the four-dimensional 
${\cal N}=1$ supersymmetric $Sp(p+M) \times Sp(p)$ gauge theory 
with two chiral fields $A^{ai\alpha}$ $(\alpha=1,2)$ in the bifundamental 
representation, four $Q^{aA}$ $(A=1,...,4)$ 
in the fundamental representation of $Sp(p+M)$ gauge group 
and four $q^{iI}$ $(I=1,...,4)$ in the fundamental representation 
of $Sp(p)$ gauge group. 
Here, $a=1,...,2(p+M)$ denotes the index of 
the $Sp(p+M)$ gauge group 
and $i=1,...,2p$ denotes the one of the $Sp(p)$ gauge group. 
This gauge theory lives on the world volume of 
$p$ D3-branes and $M$ fractional D3-branes at the tip of the 
orientifolded conifold with four D7-branes on top of an orientifold plane. 
The tree level superpotential of the model is given by
\begin{align}
W_{\mathrm{tree}} 
&= - h \left( \frac{1}{2}J_{ab} J_{cd} J_{jk} J_{li} 
\varepsilon_{\alpha\gamma} \varepsilon_{\beta\delta}
A^{ai\alpha} A^{bj\beta} A^{ck\gamma} A^{dl\delta} \right. \notag \\
& + J_{ab} J_{cd} J_{ij} \varepsilon_{\alpha\beta}
Q^{aA} A^{bi\alpha} A^{cj\beta} Q^{dA}  
 - J_{ab} J_{ij} J_{kl} \varepsilon_{\alpha\beta}
 q^{iI} A^{aj\alpha} A^{bk\beta} q^{lI} \notag \\
& - \frac{1}{2} \left. J_{ab} J_{cd} Q^{aA} Q^{bB} Q^{cB} Q^{dA}
+ \frac{1}{2} J_{ij} J_{kl} q^{iI} q^{jJ} q^{kJ} q^{lI}
\right),
\label{Wtree}
\end{align}
where $J_{ab}$ and $J_{ij}$ are the invariant antisymmetric tensors 
of the $Sp(p+M)$ and the $Sp(p)$ gauge groups, respectively.
In this note, for the matrix $J$, we will use the convention 
\begin{align}
J = i\sigma_2 \otimes {\bf 1} 
= \left( 
\begin{array}{ccccc}
 & 1 & & &  \\
-1 & & & &  \\
 & &\ddots & &  \\
 & & & &1\\
 & & &-1&
\end{array}
\right).
\end{align}
We list the symmetries of the gauge theory in the following table:
\begin{align}
\begin{array}{c||c|c||c|c|c}
 &Sp(p+M) & Sp(p) & SU(2) & SO(4)_1 & SO(4)_2 \\
 &a,b,\cdots & i,j,\cdots & \alpha,\beta,\cdots & A,B,\cdots & I,J,\cdots \\
 \hline
A^{ai\alpha} & \tiny\yng(1) & \tiny\yng(1) & \tiny\yng(1) & \mathbf{1} 
& \mathbf{1}\\
Q^{aA} & \tiny\yng(1) & \mathbf{1} & \mathbf{1} & \tiny\yng(1) & \mathbf{1} \\
q^{iI} & \mathbf{1} & \tiny\yng(1) & \mathbf{1} & \mathbf{1} & \tiny\yng(1) \\
\end{array}
\end{align}


\subsection{The Classical Moduli Space}

In order to compare the deformation parameter of the 
supergravity background with the one of the quantum constraint 
 on the corresponding branches of the gauge theory, we study 
the branches which can be regarded as the motion of D3-branes probing 
the orientifolded conifold. 
We will see that 
the corresponding branch in the classical moduli space 
is specified by
\begin{align}
&A^{\alpha=1} = \left( 
\begin{array}{cccccccccc}
a_1    &      &        &     &        \\
       & b_1  &        &     &        \\
       &      & \ddots &     &        \\
       &      &        & a_p &        \\
       &      &        &     & b_p    \\
 0     &      & \cdots &     &  0     \\
\vdots &      &        &     & \vdots \\
 0     &      & \cdots &     &  0 
\end{array}
\right), 
\qquad 
A^{\alpha=2} = \left( 
\begin{array}{cccccccccc}
c_1    &      &        &     &        \\
       & d_1  &        &     &        \\
       &      & \ddots &     &        \\
       &      &        & c_p &        \\
       &      &        &     & d_p    \\
 0     &      & \cdots &     &  0     \\
\vdots &      &        &     & \vdots \\
 0     &      & \cdots &     &  0 
\end{array}
\right) ,
\notag\\
&~~~~Q = 
\left(
\begin{array}{cccccccccc}
0 & 0 & 0  & 0 \\
\vdots & \vdots& \vdots& \vdots  \\
0 & 0 & 0 & 0 \\
a & 0 & b & c \\
0 & a & c & -b \\
0 & 0 & 0 & 0 \\
\vdots & \vdots& \vdots& \vdots  \\
0 & 0 & 0 & 0 
\end{array}
\right),
\begin{array}{l}
\left. 
\begin{array}{c}
\!\!\!\! \\
\!\!\!\! \\
\!\!\!\! 
\end{array}
\right\} 2p \\
\begin{array}{c}
\!\!\!\! \\
\!\!\!\! 
\end{array} \\
\left. 
\begin{array}{c}
\!\!\!\! \\
\!\!\!\! \\
\!\!\!\!
\end{array}
\right\} 2(M-1)
\end{array} 
q=0,
\label{vevAQ}
\end{align}
where $|a_i|^2 + |c_i|^2 = |b_i|^2 + |d_i|^2$,~ ($i=1,...,p$) and 
$a^2=b^2+c^2$.
We can verify that the vacuum expectation values (\ref{vevAQ}) satisfy 
the $F$-term conditions and the $D$-term conditions. 
We note that at a generic 
point on the moduli space, 
the $Sp(p+M)\times Sp(p)$ gauge group are broken to 
$Sp(M-1)\times U(1)^p$. 

The classical moduli space (\ref{vevAQ}) can also be given in terms of the 
chiral fields 
\begin{align}
N^{ij\alpha\beta} = J_{ab} A^{ai\alpha} A^{bj\beta} , 
\qquad
v^{i\alpha A} = J_{ab} A^{ai\alpha} Q^{bA} , 
\qquad
M^{AB} = J_{ab} Q^{aA} Q^{bB} \label{def_N},
\end{align}
which are left invariant only by the $Sp(p+M)$ gauge transformation. 
For the purpose of our discussion, 
it is convenient to introduce 
the matrix notation for the meson $N^{ij\alpha\beta}$ 
as $(\tilde{N}^{\alpha\beta})^i{}_k \equiv J_{kj} N^{ij\alpha\beta}$. 
Now the moduli space can be rewritten as 
\begin{align}
&\sqrt{2h}\tilde{N}^{11} 
= \mathrm{diag} \left( x_1, x_1, \cdots, x_p, x_p \right), 
\qquad
\sqrt{2h}\tilde{N}^{12} 
= \mathrm{diag} \left( w_1, z_1, \cdots, w_p, z_p \right),
\notag\\
&\sqrt{2h}\tilde{N}^{21} 
= \mathrm{diag} \left( z_1, w_1, \cdots, z_p, w_p \right),
\qquad
\sqrt{2h}\tilde{N}^{22} 
= \mathrm{diag} \left( y_1, y_1, \cdots, y_p, y_p \right),
\notag\\
&\sqrt{2h} M = 
\left( 
\begin{array}{cccc}
0 & X & Y & Z \\
-X & 0 & Z & -Y \\
-Y & -Z & 0 & X \\
-Z & Y & -X & 0
\end{array}
\right), 
\qquad 
v=0. 
\label{vevM}
\end{align}
The components of matrices are related to 
the vacuum expectation values (\ref{vevAQ}) as 
\begin{align}
&x_i = \sqrt{2h}a_ib_i, \quad y_i = \sqrt{2h}c_id_i, \quad 
w_i= \sqrt{2h}b_ic_i, \quad z_i= \sqrt{2h}a_id_i, 
\notag\\
&X=\sqrt{2h}a^2=\sqrt{2h}(b^2+c^2), \quad 
Y=\sqrt{2h}ac, \quad Z=-\sqrt{2h}ab, 
\end{align}
which satisfy the constraints 
\begin{align}
x_iy_i-w_iz_i=0, \qquad
X^2+Y^2+Z^2=0.
\label{classconst}
\end{align}
We introduced the factor $\sqrt{2h}$ 
in order to absorb the coupling constant $h$ 
in the superpotential (\ref{Wtree}).
In addition to the first equation of (\ref{classconst}) giving 
the singular conifold, 
there exists the $Sp(p)$ gauge transformation 
\begin{align}
(x_i , y_i , z_i, w_i ) \leftrightarrow (x_i , y_i , w_i , z_i ). 
\label{changexyzw1}
\end{align}
This gauge transformation leads to the $\mathbb{Z}_2$ identification 
on the conifold. 
Thus, the classical moduli space (\ref{vevM}) describes 
$p$ D3-branes probing the orientifolded conifold \cite{NSW}.

At least at a generic point 
far away from the origin of the classical moduli space, 
the vacuum expectation value (\ref{vevAQ}) may give a good description 
to the quantum moduli space. In particular, the gauge group 
$Sp(p+M)\times{}Sp(p)$ are broken to $Sp(M-1)\times U(1)^p$ 
at the point, where the effective theory 
is given by the $Sp(M-1)$ pure Yang-Mills theory. 
The classical $U(1)_R$ symmetry 
of the this pure Yang-Mills theory is broken by 
the anomaly to $\mathbb{Z}_{2M}$ symmetry, which rotates the gaugino as 
\begin{align}
\lambda\to\lambda e^{i\frac{2\pi}{2M}}. 
\end{align}
Furthermore, the $\mathbb{Z}_{2M}$ symmetry is broken by gaugino condensation 
to $\mathbb{Z}_{2}$, and we end up with $M$ inequivalent vacua in the infrared.


\subsection{The Quantum Moduli Space}

In this section, we will see that the classical moduli space 
is deformed by the quantum effect. 
In the case for $p\ge M+1$, 
we can perform the Seiberg duality transformation 
\cite{EM} at least once, and in general several times. 
After performing the transformations say, 
$l$ times and integrating out the singlets 
under the $Sp(p-(l-1)M)$ gauge group, 
we can find a branch where the mesons of the dual quarks develop 
the vacuum expectation values. 
On the branch, we will see that the classical constraints of 
the mesons, describing the conifold, 
are quantum-mechanically deformed 
to the one describing the deformed conifold 
with the deformation parameter 
given by the dynamical scale, as in the case $p=1$ \cite{NSW}. 
As found in \cite{DKS}, 
we will find the deformation parameters on the meson 
branches depend on the numbers $l$ of performing 
the duality transformations. 
We will demonstrate the ratio of two different deformation parameters 
matches the results on the gravity side in the previous section.

Taking into account the 1-loop $\beta$-functions 
of the $Sp(p+M)$ and $Sp(p)$ gauge groups, 
the gauge coupling constant of $Sp(p+M)$ 
becomes much larger than the one of $Sp(p)$ 
after a finite amount of RG flow. 
Therefore, we will effectively regard the theory as $Sp(p+M)$ 
gauge theory with $N_f=2p+2$ flavors. 
We denotes the dynamical scale of the $Sp(p+M)$ gauge theory 
as $\Lambda_1$ and the one of $SU(p)$ as $\Lambda_2$. 
Since at the energy scale lower than $\Lambda_1$ 
the magnetic theory gives a good description, 
we perform the Seiberg duality transformation \cite{EM} 
to go to the dual 
$Sp(p-M) \times Sp(p)$ gauge theory.  
The matter fields in the dual theory are listed in the table:
\begin{align}
\begin{array}{c||c|c||c|c|c}
 &Sp(p-M) & Sp(p) & SU(2) & SO(4)_1 & SO(4)_2 \\
 &\tilde{a},\tilde{b},\cdots & i,j,\cdots & \alpha,\beta,\cdots 
 & A,B,\cdots & I,J,\cdots \\
 \hline
\tilde{A}^{\tilde{a}}{}_{i\alpha} & \tiny\yng(1) & \tiny\yng(1) 
& \tiny\yng(1) & \mathbf{1} & \mathbf{1}\\
\tilde{Q}^{\tilde{a}}{}_A & \tiny\yng(1) & \mathbf{1} & \mathbf{1} 
& \tiny\yng(1) & \mathbf{1} \\
q^{iI} & \mathbf{1} & \tiny\yng(1) & \mathbf{1} & \mathbf{1} & \tiny\yng(1) \\
\hat{N}^{[ij](\alpha\beta)} & \mathbf{1} & \tiny\yng(1,1) & \tiny\yng(2) 
& \mathbf{1} & \mathbf{1}\\
n^{(\alpha\beta)} & \mathbf{1} & \mathbf{1} & \tiny\yng(2) & \mathbf{1} 
& \mathbf{1}\\
N^{(ij)[\alpha\beta]} & \mathbf{1} & \tiny\yng(2) & \mathbf{1} 
& \mathbf{1} & \mathbf{1}\\
v^{i\alpha A}  & \mathbf{1} & \tiny\yng(1) & \tiny\yng(1)  
& \tiny\yng(1) & \mathbf{1} \\
M^{AB} & \mathbf{1} & \mathbf{1} & \mathbf{1} & \tiny\yng(1,1) & \mathbf{1} 
\end{array}
\end{align}
Here, we decomposed the $Sp(p-M)$ singlet meson 
$N^{ij\alpha\beta}$ into the irreducible representations as 
\begin{align}
N^{ij\alpha\beta}
=N^{(ij)[\alpha\beta]}+\widetilde{N}^{[ij](\alpha\beta)}
+n^{(\alpha\beta)}J^{ij}. 
\end{align}
The superpotential of the magnetic theory is given by 
\begin{align}
&W_{\mathrm{tree}} =
-h\left( 
\frac{1}{2} J_{jk}J_{li} \varepsilon_{\alpha\gamma} 
\varepsilon_{\beta\delta} N^{ij\alpha\beta}N^{kl\gamma\delta}
- J_{ij} \varepsilon_{\alpha\beta} v^{i\alpha A}v^{j\beta A} \right. 
\notag\\
&\hspace{2cm} \left.
- J_{jk}J_{li} \varepsilon_{\alpha\beta}N^{ij\alpha\beta}q^{kI}q^{lI}
- \frac{1}{2}M^{AB}M^{BA} 
+ \frac{1}{2}J_{ij}J_{kl} q^{iI}q^{jJ}q^{kJ}q^{lI} \right) 
\notag \\
&\hspace{1.5cm}
+ \frac{1}{\mu}
\left( J_{\tilde{a}\tilde{b}} 
N^{ij\alpha\beta} \tilde{A}^{\tilde{a}}{}_{i\alpha} 
\tilde{A}^{\tilde{b}}{}_{j\beta}
+ 2J_{\tilde{a}\tilde{b}}
v^{i\alpha A} \tilde{A}^{\tilde{a}}{}_{i\alpha} \tilde{Q}^{\tilde{b}}{}_A 
+ J_{\tilde{a}\tilde{b}} M^{AB} 
\tilde{Q}^{\tilde{a}}{}_A \tilde{Q}^{\tilde{b}}{}_B\right), 
\label{magsuperpot}
\end{align}
where $\mu$ is the duality scale. 


We can see that the superpotential (\ref{magsuperpot}) gives the mesons 
$N$, $v$, and $M$ masses of the order $h|\Lambda_1|^2$. 
On the other hand, 
when the fields $N$ and $M$ develop vacuum expectation values, 
$\tilde{A}$ and $\tilde{Q}$ acquire the masses 
${\langle N \rangle/\mu}$ and ${\langle M \rangle}/{\mu}$
via the superpotential (\ref{magsuperpot}). 
Therefore, in order to obtain the correct low energy effective theory 
by integrating out massive degrees of freedom, 
we need to consider the case 
$\left| \langle N \rangle/{\mu} \right|,
\left| \langle M \rangle/{\mu}\right| \gg \left| h\Lambda_1^2 \right|$ 
and the case 
$\left| \langle N \rangle/{\mu} \right|,
\left| \langle M \rangle/{\mu} \right| \ll \left|h\Lambda_1^2 \right|$, 
separately. 

We begin with the former case 
${\left| \langle N \rangle/{\mu} \right|},
{\left| \langle M \rangle/{\mu} \right|}\gg \left|h\Lambda_1^2\right|$, 
where the mesons $N$ and $M$ obtain vacuum expectation 
values of full rank and its eigenvalues are all much larger than 
$h\mu\Lambda_1^2$. Since the dual quarks $\tilde{A},\tilde{Q}$ acquire 
masses much larger than the mesons $N$, $v$ and $M$, 
we integrate them out to obtain the pure $Sp(p-M)$ gauge theory. 
The gaugino condensation generates the superpotential  
\begin{align}
W_{{\rm dyn}}=(p-M+1) \Lambda_L^3,
\label{gaugino}
\end{align}
where $\Lambda_L$ is the dynamical scale of 
the pure super Yang-Mills theory.

In order to obtain the low energy effective potential for 
the mesons $N$ and $M$ from the superpotential (\ref{gaugino}), 
it is convenient to write the mesons 
as a matrix 
\begin{align}
V=(V^{zw})=\left(
\begin{array}{cc}
N^{ij\alpha\beta} & v^{j\beta{B}} \\
-v^{i\alpha{A}} & M^{AB}
\end{array}
\right),
\label{def_V}
\end{align}
where $z$, $w$ run over both of the paired indices $(i,\alpha)$ of the 
$Sp(p)$ gauge group and the global $SU(2)$ group and the index $A$ of 
the global $SO(4)$ group.
We have a matching condition of the dynamical scales
\begin{align}
\Lambda_L^{3(p-M+1)} 
= \tilde{\Lambda}_2{}^{p-3M+1}
\left(\mathrm{Pf}\langle V \rangle \over \mu^{2p+2}\right), 
\label{matchingcond1}
\end{align}
where $\tilde{\Lambda}_2$ is the dynamical scale 
of the magnetic $Sp(p-M)$ theory 
with the dual quarks $\tilde{A}$ and $\tilde{Q}$. 
We also have a relation between the dynamical scales 
$\Lambda_1$ and $\tilde{\Lambda}_2$ 
\begin{align}
\Lambda_1{}^{p+3M+1} \tilde{\Lambda}_2{}^{p-3M+1} = (-1)^{p-M+1}\mu^{2p+2}. 
\label{duality_relation}
\end{align}
Using the relations (\ref{matchingcond1}) and (\ref{duality_relation}),  the superpotential (\ref{gaugino}) can be written as  
\begin{align}
W_{{\rm dyn}}=-(p-M+1)\left(\frac{\mathrm{Pf}\left(V\right) }
{{\Lambda_1}^{p+3M+1}}\right)^{\frac{1}{p-M+1}}. 
\label{ADS_pot}
\end{align}
Thus, we find the total superpotential is given by
\begin{align}
&W_{{\rm eff}} =
-h\left( 
\frac{1}{2} J_{jk}J_{li} \varepsilon_{\alpha\gamma} 
\varepsilon_{\beta\delta} N^{ij\alpha\beta}N^{kl\gamma\delta}
- J_{ij} \varepsilon_{\alpha\beta} v^{i\alpha A}v^{j\beta A}
- J_{jk}J_{li} \varepsilon_{\alpha\beta} N^{ij\alpha\beta}q^{kI}q^{lI} 
 \right. 
\notag\\
&\hspace{1.8cm}
\left.
- \frac{1}{2} M^{AB}M^{BA}
+ \frac{1}{2} J_{ij}J_{kl} q^{iI}q^{jJ}q^{kJ}q^{lI} \right) 
- (p-M+1)\left(\frac{\mathrm{Pf}\left(V\right)}
{{\Lambda_1}^{p+3M+1}}\right)^{\frac{1}{p-M+1}}.
\label{Weff0}
\end{align}
We substitute the vacuum expectation values (\ref{vevM}) 
into the superpotential (\ref{Weff0}) and 
solve the $F$-term conditions. 
We find that the classical constraints (\ref{classconst}) are deformed into 
\begin{align}
x_i y_i - z_i w_i = \varepsilon^2_{l=0,r}, ~~(i= 1,\cdots, p ), 
\qquad 
X^2+Y^2+Z^2 = \varepsilon^2_{l=0,r}, \label{quantum_moduli}
\end{align}
where the deformation parameter $\varepsilon_{l=0,r}$ is given by
\begin{align}
\varepsilon^2_{l=0,r}
=\left[(2h)^{p+1} \Lambda_1{}^{p+3M+1}\right]^{\frac{1}{M}}
e^{{2\pi\over{M}}ir}\qquad (r=0,...,M-1). 
\label{dp0}
\end{align} 
We refer this branch as the $l=0$ branch. 
The deformed constraint corresponds to the motion of 
$p$ D3 branes probing the deformed conifold. 
The phase factor $e^{({2\pi/{M}})ir}$ comes from the branch cut in 
the deformation parameter $\varepsilon_{l=0,r}$ 
and thus gives additional $M$ distinct 
branches. 
These $M$ branches are generated by the symmetry breaking 
from $\mathbb{Z}_{2M}$ to $\mathbb{Z}_2$ 
through gaugino condensation, as discussed.

Now, let us turn to the other case $\left| \langle N \rangle/\mu \right|$, 
$\left| \langle M \rangle/\mu \right| \ll \left| h\Lambda_1^2 \right|$. 
In this case, 
the mesons $N$ and $M$ are much heavier than the dual quarks $\tilde{A}$ 
and $\tilde{Q}$, and we integrate out the mesons. 
We obtain the 
$Sp(p) \times Sp(p-M)$ theory with matters listed as follows: 
\begin{align}
\begin{array}{c||c|c||c|c|c}
 &Sp(p) & Sp(p-M) & SU(2) & SO(4)_2 & SO(4)_1\\
 &i,j,\cdots & \tilde{a},\tilde{b},\cdots  & \alpha,\beta,\cdots 
 & I,J,\cdots & A,B,\cdots \\
 \hline
\tilde{A}^{i\tilde{a}\alpha} 
& \tiny\yng(1) & \tiny\yng(1) & \tiny\yng(1) & \mathbf{1} & \mathbf{1} \\
q^{iI} & \tiny\yng(1) & \mathbf{1} & \mathbf{1} & \tiny\yng(1) & \mathbf{1} \\
\tilde{Q}^{\tilde{a}A}& \mathbf{1} & \tiny\yng(1) & \mathbf{1} & \mathbf{1} 
& \tiny\yng(1) \\
\end{array}
\end{align}
The superpotential is given by 
\begin{align}
\widetilde{W}_{{\rm tree}}
&= -\frac{1}{h\mu^2} \left( \frac{1}{2} J_{jk} J_{li} 
J_{\tilde{a}\tilde{b}} J_{\tilde{c}\tilde{d}} 
\varepsilon_{\alpha\gamma} \varepsilon_{\beta\delta}
\tilde{A}^{i\tilde{a}\alpha}
\tilde{A}^{j\tilde{b}\beta}
\tilde{A}^{k\tilde{c}\gamma}
\tilde{A}^{l\tilde{d}\delta} \right. \notag \\
& + h\mu J_{ij} J_{kl}
J_{\tilde{a}\tilde{b}} \varepsilon_{\alpha\beta}
q ^{iI} \tilde{A}^{j\tilde{a}\alpha} \tilde{A}^{k\tilde{b}\beta} q^{lI} 
- J_{ij} J_{\tilde{a}\tilde{b}} J_{\tilde{c}\tilde{d}} 
 \varepsilon_{\alpha\beta}
 \tilde{Q}^{\tilde{a}A} \tilde{A}^{i\tilde{b}\alpha} 
 \tilde{A}^{j\tilde{c}\beta}\tilde{Q}^{\tilde{d}A}  \notag \\
&  - \left. \frac{(h\mu)^2}{2} J_{ij}J_{kl} q^{iI}q^{jJ}q^{kJ}q^{lI}
+ \frac{1}{2} 
J_{\tilde{a}\tilde{b}} J_{\tilde{c}\tilde{d}}
\tilde{Q}^{\tilde{a}A} \tilde{Q}^{\tilde{b}B}
\tilde{Q}^{\tilde{c}B} \tilde{Q}^{\tilde{d}A}
\right) .
\label{magWtree}
\end{align}
The resulting theory is similar to the original electric theory, 
if we make the replacement 
$p \to p-M$, $A \to \tilde{A}$, $Q \to q$ and $q \to \tilde{Q}$ 
in the original theory. 
We obtain a new quantum constraints 
in terms of the mesons 
$\widetilde{N}$, $\widetilde{M}$, $\widetilde{v}$ 
of the dual quarks $\widetilde{A},\widetilde{Q}$ 
and the electric quark $q$ in the form (\ref{vevM}). 
We refer this branch as the $l=1$ branch. 

Now, let us calculate the deformation parameter of the $l=1$ branch. 
We can borrow the result of the $l=0$ branch. 
In order to this, 
we must introduce the scale factor $\sqrt{2/(h\mu^2)}$ 
for the meson $\widetilde{N},\widetilde{v}$ and 
$2h$ for $\widetilde{M}$ in the similar way as (\ref{vevM}) 
and use the dynamical scale $\hat{\Lambda}_1$ 
of the $Sp(p)$ gauge theory. 
We find the deformation parameter $\varepsilon_{l=1,r}$ is  
\begin{align}
\varepsilon^2_{l=1,r} = 
\left[2^{p-M+1} h^{-p+M+1}\mu^{2(-p+M)}{\hat{\Lambda}}_1{}^{p+2M+1}
\right]^{\frac{1}{M}} e^{{2\pi\over{M}}ir}\qquad (r=0,...,M-1). 
\end{align}
The phase factor $e^{{2\pi\over{M}}ir}$ has the same origin 
as the one for the $l=0$ branch, 
and there exist $M$ branches.

Now let us recall that we can perform 
the duality transformations successively. 
When we perform the duality transformations $l_0$ times, 
we find the ``$l=0$" branch 
in the $Sp(p-(l_0-1)M)\times Sp(p-l_0M)$ gauge theory, 
which we refer as the $l=l_0$ branch 
of the $Sp(p+M) \times Sp(p)$ gauge theory. 
Similarly to our discussion in the previous section, 
the branches of $l=0,1,\cdots ,k$ may be regarded as $p-lM$ D3-brane 
probes on the orientifolded deformed conifold.

The maximum number of times 
that we can perform the duality transformations is 
$k=\left[{p}/{M}\right]$. 
At the end of the duality cascade 
we have $Sp(p' + M) \times Sp(p')$ gauge theory, where $p'=p-kM$.  
We can effectively regard $Sp(p'+M)$ gauge theory with $N_f=2p'+2$ flavors. We find $N_f\le N_c+1$ at the end of the duality cascade. 
In the case for $p'\le M-2~(N_f\le N_c-1)$, 
the dynamics of the $Sp(p^{\prime} + M)$ gauge theory 
generates the non-perturbative superpotential \cite{IP}
\begin{align}
(-p^{\prime}+M-1) \left(\frac{{\Lambda_1}^{p^{\prime}+3M+1}}{\mathrm{Pf} V}
\right) ^{\frac{1}{-p^{\prime}+M-1}},
\end{align}
on the branch in the form (\ref{vevM}) to quantum-mechanically modify the 
classical constraints (\ref{classconst}) into the one for the deformed 
conifold. The case of $p'=M-1, M~(N_f=N_c,N_c+1)$ 
leads to the similar result and is discussed in the appendix. 

Now let us evaluate the ratio of the deformation parameters 
$\varepsilon_{l=1}$ and $\varepsilon_{l=0}$. 
For this purpose, we would like to write the deformation parameter 
$\varepsilon_{l=1}$ 
in terms of the dynamical scales of the original $Sp(p+M)\times{}Sp(p)$ 
gauge theory. 
We have the anomalous symmetries which rotate 
the fields, the coupling constant and the instanton factors: 
\begin{align}
\begin{array}{c|c|c}
& U(1)_A & U(1)_R \\
\hline 
A,Q,q & 1 & 1 \\
h & -4 & -2 \\
\Lambda_1{}^{p+3M+1} & 4p+4 & 2(p+M)+2 \\
\Lambda_2{}^{p-2M+1} & 4(p+M)+4 & 2p+2
\end{array}
\end{align}
The ratio of the two deformation parameters 
should be the function of the quantity 
which is 
invariant under these symmetries as well as dimensionless. 
Such a quantity 
turns out to be
\begin{align}
h^{2p+M+2} \Lambda_1{}^{p+3M+1} \Lambda_2{}^{p-2M+1} 
\label{ratio}. 
\end{align}

We may assume that the scale $\mu$ appearing in the magnetic superpotential 
(\ref{magsuperpot}) provides the cut-off below which the magnetic theory 
gives a good description for the low energy dynamics of the electric theory. 
At energies close to the scale $\mu$, it is conceivable that the coupling 
constant $\Lambda_2$ of the $Sp(p)$ gauge dynamics also matches 
the one $\tilde\Lambda_1$ of the $Sp(p)$ after the duality transformation as 
\begin{align}
\left|\,{\tilde{\Lambda}_1\over\mu}\,\right|^{-3p+2M-1} \sim\qquad 
\left|{\Lambda_2\over\mu}\right|^{p-2M+1}.
\end{align}
The mesons $N$, $v$ and $M$ are integrated out 
to obtain the $l=1$ branch, 
and the matching condition at the energy scale of their mass 
$h|\Lambda_1|^2$ gives 
\begin{align}
{\hat{\Lambda}}_1 {}^{p+2M+1} \sim 
\tilde{\Lambda}_1{}^{-3p+2M-1} 
(h|\Lambda_1|{}^2)^{4p+2}. 
\end{align}
As in \cite{DKS}, we follow the convention
\begin{align}
|\mu| \sim |\Lambda_1|. 
\end{align}

Using these relations, 
together with the relation between $\Lambda_1$ and 
$\tilde{\Lambda}_2$ in (\ref{duality_relation}), 
we find the 
deformation parameter $\varepsilon_{l=1}$ is
\begin{align}
\left| \varepsilon^2_{l=1} \right|
&\sim \left| \left[ h ^{3p+M+3} \Lambda_1{}^{2p+6M+2} \Lambda_2{}^{p-2M+1} 
\right] ^{\frac{1}{M}} \right| \notag \\
&\sim \left| \varepsilon_{l=0}^2 \left[
h ^{2p+M+2} \Lambda_1{}^{p+3M+1} \Lambda_2{}^{p-2M+1} \right] ^{\frac{1}{M}} 
\right|.
\end{align}
Since the ratio should be given in terms of the combination (\ref{ratio}), 
we find that the ratio is given by 
\begin{align}
\frac{\varepsilon^2_{l=1}}{\varepsilon^2_{l=0}}
\sim \left[ h ^{2p+M+2} \Lambda_1{}^{p+3M+1} 
\Lambda_2{}^{p-2M+1} \right] ^{\frac{1}{M}}.
\label{result}
\end{align}

The relation of the string coupling with 
the gauge coupling constant $g^2_1$ of the $SU(p+M)$ gauge group 
and the coupling $g^2_2$ of the $SU(p)$ is known as \cite{KW,MP}
\begin{align}
{8\pi^2\over{g}^2_1}+{8\pi^2\over{g}^2_2}\sim{2\pi\over{g}_s}.
\end{align}
Since the $Sp(N)$ gauge coupling constant $1/g^2_{Sp}$ on the worldvolume theory 
is half as large as the $SU(N)$ gauge coupling constant $1/g^2_{SU}$, 
as in \cite{DKS}, it leads us to the relation 
\begin{align}
\exp\left(-{2\pi\over 2{g}_s}\right)
\sim h ^{2p+M+2} \Lambda_1{}^{p+3M+1} \Lambda_2{}^{p-2M+1}.
\end{align}
Using this relation, 
we find that the result (\ref{result}) exactly matches 
with the result (\ref{stringresult}) on the gravity side 
discussed in the previous section.


\section{Summary and Discussion}

We have studied the quantum moduli space of 
the cascading $Sp(p+M)\times Sp(p)$ gauge theory. 
We found that the gauge theory includes many branches 
labeled by $r=0,...,M-1$ and $l=0,...,[p/M]$. 
The label $r$ distinguishes $M$ vacua 
generated by the symmetry breaking through gaugino condensation,
while the label $l$ is the number of times we take 
the Seiberg duality transformations 
and specifies the dimension of the flat directions.
The branch $l$ is proposed in \cite{DKS} to correspond to the KS solution 
with mobile $(p-lM)$ D3-branes for the $SU(p+M)\times{}SU(p)$ gauge theory. 
In this note, we applied their proposal to the $Sp(p+M)\times{}Sp(p)$ 
gauge theory and have seen that the ratio of the deformation parameters of 
different two branches is in exact agreement with the result on the string 
dual. 

It would be an interesting problem concerned with the 
orientifolded conifolds to study the world volume theory on 
the $p$ D3-branes and $M$ fractional D3-branes 
at the orientifolded conifold without D7-branes. 
The gauge theory living on the D3-branes 
has only chiral bifundamental fields and no fundamental fields. 
The beta function and the symmetries of the gauge theory 
are different from the ones we discussed. 
In particular, as found in \cite{ouyang} for the $SU(p+M)\times{}SU(p)$ 
cascade gauge theory, at each step of the cascade duality 
transformation, the difference between the ranks of the one factor 
group and the one of the other of the gauge groups change. 
On the other hand, on the supergravity side, 
the axion-dilaton field, which couples to the O7-plane, is no longer 
constant,  since we have no D7-branes to cancel the RR-charge of the O7-plane. 
The axion-dilaton field thus gets a monodromy around the orientifold 
fixed point. 
Furthermore, it turns out that the number of RR 3-form flux as well as 
RR 5-form flux has the radial dependence. 
Therefore, it would be interesting to construct the concrete supergravity 
solution and analyze the gauge theory via the holographic correspondence.


\section*{Acknowledgements}
We would like to thank Masahiro Ibe, Yu Nakayama, Yuji Tachikawa 
and especially to Teruhiko Kawano and Yosuke Imamura
for useful discussions. 
The research of F. K. was supported in part by the JSPS predoctoral fellowship. 


\section*{Appendix}

\appendix

\section{Quantum moduli space for $p=M-1,M$}

In this appendix, we discuss that the classical moduli space for $p=M-1,M$ 
are also modified by the quantum effect to give rise to 
the constraint for the deformed conifold, in a similar way to the case 
for $p < M-1$.  

In the case $p=M-1$, the non-perturbative superpotential
\begin{align}
W_{{\rm dyn}} = X \left( \mathrm{Pf} (V) - {\Lambda_1}^{p+3M+1} \right),
\end{align}
where $X$ is a Lagrange multiplier, 
is generated \cite{IP} to give the low energy effective superpotential 
\begin{eqnarray}
&&W_{\mathrm{eff}}= h\left( 
J_{jk}J_{li} \varepsilon_{\alpha\gamma} \varepsilon_{\beta\delta} 
	N^{ij\alpha\beta}N^{kl\gamma\delta}
+ J_{ij} \varepsilon_{\alpha\beta} v^{i\alpha A}v^{j\beta A} 
+ M^{AB}M^{BA} \right. 
\nonumber\\
&&\hspace{1.7cm}
+ J_{jk}J_{li} \varepsilon_{\alpha\beta} 
N^{ij\alpha\beta}q^{kI}q^{lI} 
+ J_{ij}J_{kl} q^{iI}q^{jJ}q^{kJ}q^{lI} ) 
\\
&&\hspace{1.7cm}
+  X \left( \mathrm{Pf} (V) - {\Lambda_1}^{p+3M+1} \right).
\nonumber
\end{eqnarray}
On the branch in the form (\ref{vevM}), 
solving the $F$-term conditions from the effective superpotential, 
we obtain the quantum constraint (\ref{quantum_moduli}), giving 
the deformed conifold.

Also for $p=M$, the non-perturbative dynamics generates 
the superpotential \cite{IP}
\begin{align}
W_{{\rm dyn}} = -\frac{\mathrm{Pf} (V)}{\Lambda_1{}^{p+3M+1}}.
\end{align}
In order to find the quantum moduli space on the branch (\ref{vevM}), 
since the superpotential can be obtained formally by substituting $p=M$ into 
(\ref{ADS_pot}), we only have to solve the same $F$-term condition 
as there but with $p=M$ and find the quantum deformed constraint 
(\ref{quantum_moduli}).


\end{document}